\documentclass[prb,aps,showpacs,twocolumn,unsortedaddress,longbibliography,superscriptaddress]{revtex4-1}
\usepackage{latexsym}
\usepackage{graphicx}
\usepackage{color}
\usepackage{amsmath}
\usepackage{amssymb}
\usepackage{soul}

\setcitestyle{super,open={},close={}}

\makeatletter
\renewcommand\@biblabel[1]{#1.}
\makeatother

\begin{document}
\title{Bulk-dissociated topological bands without spin-orbit coupling in hetero-dimensional superconducting metamaterials}

\author{J.J.~Cuozzo}\email{jjcuozzo@utep.edu}\affiliation{Department of Physics, The University of Texas at El Paso, El Paso, TX, USA.}
\author{S.A.A.~Ghorashi}\affiliation{Department of Chemistry, University of Pennsylvania, Philadelphia, USA.}
\author{P.P.~Paudel}\affiliation{Department of Physics, The University of Texas at El Paso, El Paso, TX, USA.}
\author{D.~Huber}\affiliation{Center for Integrated Nanotechnologies, Sandia National Laboratories, Albuquerque, New Mexico 87185, USA.}
\author{W.~Pan}\affiliation{Sandia National Laboratories, Livermore, CA 94551, USA.}
\author{F.~L\'{e}onard}\affiliation{Sandia National Laboratories, Livermore, CA 94551, USA.}

\date{\today}

\begin{abstract}
    Topological superconductors (TSCs) in superconducting hybrid heterostructures, which integrate superconducting and non-superconducting materials, have been intensely investigated with the hope of discovering exotic non-Abelian anyons for fault-tolerant quantum computing. In this effort, a challenge for hybrid superconducting systems is controlling hybridization, which is often a balance between enhancing the superconducting proximity effect at the cost of suppressing desirable electronic properties such as strong spin-orbit interactions. Hence, discovering hybrid superconducting systems with topological properties controlled and enhanced by material geometry design without spin-orbit interactions would be intriguing to explore. In this work, we theoretically study a square superconducting network decorated with spin-polarized magnetic adatoms. We find that localized Yu-Shiba-Rusinov bound states at magnetic adatom sites collectively form a weak topological superconducting phase despite the absence of spin-orbit interactions. We then demonstrate that by tuning the Fermi energy of the network, the system can transition from a weak TSC phase to a bulk-dissociated TSC phase where the edge state bands separate from the bulk, giving rise to unexpected features such as nodal lines and co-existing bulk-dissociated edge and corner modes. Moreover, our findings highlight how hetero-dimensional superconducting metamaterials can serve as a useful template for controlling the coupling and dissociation between electronic degrees of freedom of different dimensionalities.
\end{abstract}

\maketitle

\section{Introduction}
Geometric structures and patterns are ubiquitous in nature, ranging in scale from the macroscopic (e.g. zebra stripes) to the microscopic (e.g. DNA double helix). The development of precision growth and fabrication tools over the past few decades have led to the advent of metamaterials-- materials geometrically optimized for a particular function. Nowadays, metamaterials play an important role in many optical~\cite{Smith2000, Walser2001, Chen2025}, acoustic~\cite{Ma2016, Cummer2016}, and thermal~\cite{Li2021, Sklan2018} devices where lengths scales are relatively large, but generating geometric motifs at much smaller scales to modify electronic systems is a challenge limited by the resolution of fabrication tools. Recent capability developments in nano-assembly of low dimensional materials have made it possible to create nanoscale metamaterials, opening the door to realizing exotic electronic states~\cite{Cao2018, Sharpe2019, Kurebayashi2022, Lau2022, Andrei2021}. While much attention has been paid to nanoscale heterostructures using materials of the same dimensionality, we seek to explore how \textit{hetero-dimensional} nanoscale metamaterials may unlock unique advantages in superconducting materials and devices.

A particular class of electronic systems that would stand to benefit from metamaterial optimization are topological phases of matter. Topological insulating phases in solid state physics can be described by a bulk energy gap, bulk topological invariant, and robust non-local electronic phenomena e.g. gapless boundary states arising in accordance with the bulk-boundary correspondence~\cite{Hasan2010, Qi2011}. The earliest descriptions of topological phases emerged from the study of the integer quantum Hall effect decades ago, and since then the field has grown dramatically to describe a wide variety of gapped and gapless phases. 
Topological phases of matter are also promising systems to create nascent nanoscale and quantum technologies. For instance, spinless p-wave topological superconductors (TSCs) have been of particular interest owing to their use as the basis for topological fault-tolerant quantum computing~\cite{Kitaev2001, Kitaev2003, Nayak2008, Alicea2012}.
Given that naturally occurring topological superconductors are rare, one of the main approaches to realize topological superconductivity is to construct hybrid materials engineered to effectively generate a topological phase. Usually these hybrid materials are heterostructures involving a conventional superconductor and a semiconductor with strong spin-orbit interactions~\cite{Lutchyn2010, Oreg2010, Sau2010, Schiela2024}. 
By carefully controlling the interface between the two materials, the combination of conventional superconductivity, strong spin-orbit coupling and a magnetic field may give rise to a spinless p-wave TSC~\cite{Lutchyn2010, Oreg2010, Sau2010}.
In this effort, one prominent challenge has been mitigating the effects of disorder which weaken the topological advantage in quantum computing and are exacerbated by large applied magnetic fields. To address this issue, alternative mechanisms to break time-reversal symmetry, such as phase-biasing~\cite{Lesser2021}, have been proposed to create spinless p-wave TSCs.
Another challenge is that compatible conventional superconductors typically have weak spin-orbit interactions that may lead to a weak renormalized spin-orbit coupling in the hybrid structure. This is problematic since TSC phases often rely on strong spin-orbit coupling (SOC); and, further complicating the issue, SOC in superconducting hybrid heterostructures is presently not well-characterized in experiments~\cite{Schiela2024}. 
One approach to overcome this problem is to develop suitable materials with stronger spin-orbit interactions, but it would be perhaps more advantageous to alleviate this issue by uncovering topological superconducting phases requiring little or no spin-orbit interactions at all.
%

\begin{figure*}[t]
	\centering
	\includegraphics[width=0.99\linewidth]{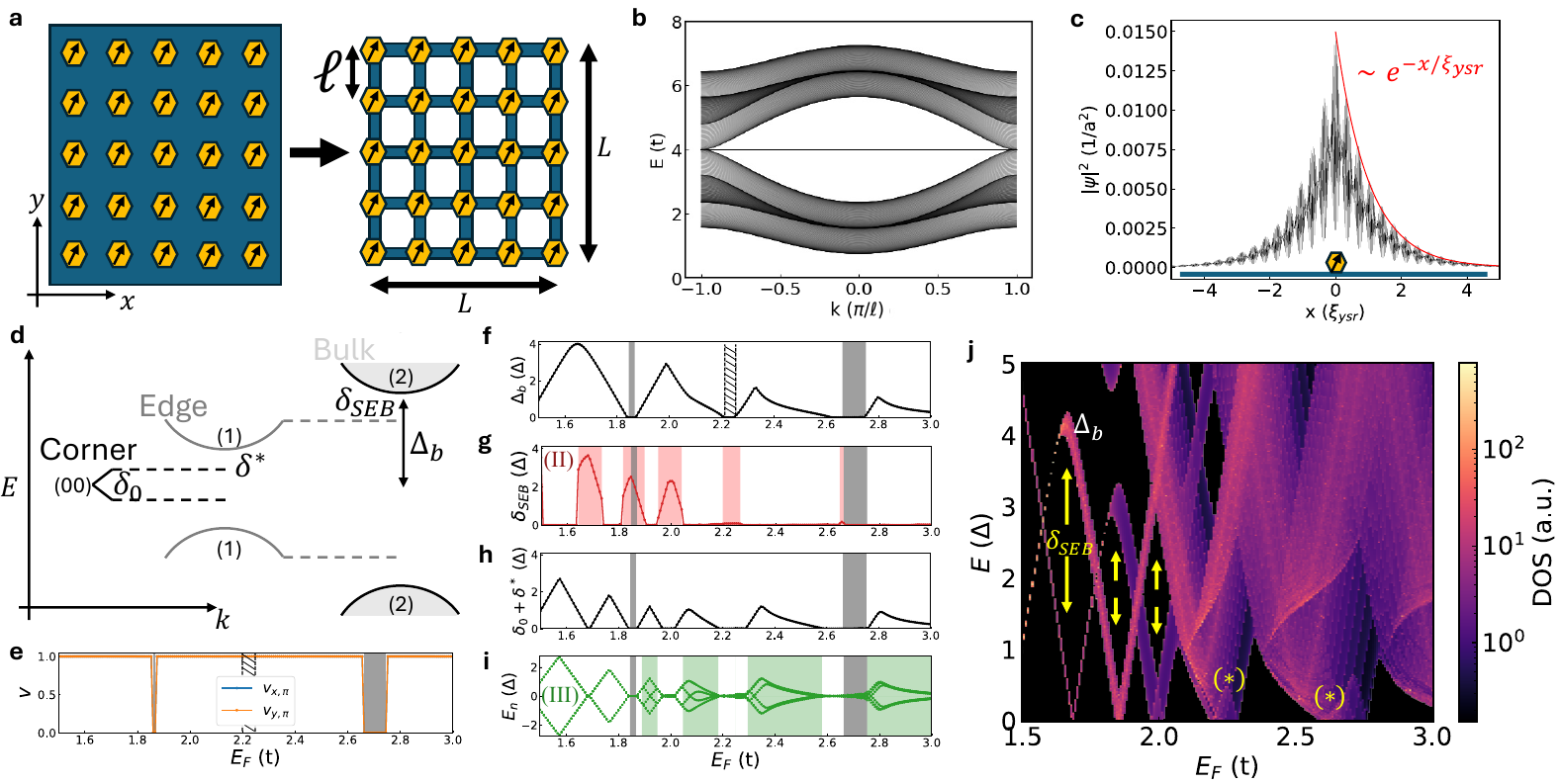} 
	\caption{\label{fig:fig1} 
		\textbf{a} A schematic of YSR square lattice (left) and a YSR square network (right) with aligned magnetic moments.
		\textbf{b} The dispersion of a translationally invariant YSR network nanoribbon in the normal state with $\ell = 10a$ and $JS = 0.6t$.
		\textbf{c} YSR bound state in 1D superconducting chain with characteristic length scale $\xi_{ysr}$.
		\textbf{d} A schematic of corner bound states and edge and bulk superconducting bands with various spectral gaps.
		\textbf{e} The weak topological invariants $v_{x/y,\pi}$, and 
        \textbf{f} spectral gaps $\Delta_b$, 
        \textbf{g} $\delta^{(direct)}_{SEB}$,
        \textbf{h} $\delta_0+\delta^*$, and 
        \textbf{i} bound state energies $E_n$ versus $E_F$.
		\textbf{j} The nanoribbon density of states versus $E_F$. Here we used $JS = 0.6t$ and $\ell = 10a$.
	}
\end{figure*}

An approach to synthesize a TSC in a hybrid structure requiring no intrinsic SOC is to couple magnetic adatoms on the surface of a superconductor, forming a so-called Yu-Shiba-Rusinov (YSR) chain~\cite{Nadj-Perge2013, Nadj-Perge2014}. Exchange coupling of nearby magnetic moments via the Ruderman–Kittel–Kasuya–Yosida (RKKY) interaction can produce a noncollinear spin texture in real space that generates a synthetic spin-orbit interaction in the chain. The synthetic SOC combined with broken TRS from the magnetic moments and superconductivity from the substrate can create a spinless p-wave TSC~\cite{Nadj-Perge2013, Nadj-Perge2014}. 
Later, this idea was generalized to two-dimensional topological phases in the form of YSR lattices where TSC and higher-order TSC phases have been predicted~\cite{Rontynen2015, Li2016_shiba}.
Two-dimensional higher-order topological superconducting phases hosting 0D topological corner modes have recently garnered interest in Shiba networks for their potential to be used in topological quantum computing~\cite{Soldini2023, Conte2024, Bruning2025, Rodriguez2025}. In this case, the corner modes are Majorana zero modes that obey many of the same properties as Majorana zero modes in one-dimensional TSCs, but they have a greater zero-energy degeneracy due to the higher dimensionality of the 2D structure.
Higher-order topological superconductors have also been studied in anti-ferromagnetic systems with a staggered flux where intrinsic spin-orbit interactions are not necessary~{\cite{Xiao2025}}.
Scanning tunneling microscopy reports on Shiba lattices indicated the possible presence of a TSC~\cite{Conte2022, Soldini2023, Bazarnik2023}, but unambiguous evidence of spinless p-wave TSCs is still lacking~\cite{Lee2012_zb_anom, Kells2012, Reeg2018, Vuik2019, Liu2019, Chen2019, Awoga2019, Woods2019, Valentini2021, Hess2021, microsoft2022, microsoft2025, Mishra2025, Pal_2024}. Furthermore, while these platforms do not require intrinsic spin-orbit interactions, they still require synthetic SOC from coupled magnetic adatoms, which is also difficult to control and characterize experimentally.

In this work, we construct a YSR superlattice by decorating a two-dimensional conventional superconducting network with magnetic impurities, see Fig.~\ref{fig:fig1}a. 
In our work, we explicitly assume collinear spin structure and no spin-orbit interactions in our model so as to describe a spin-polarized YSR superlattice. Despite this, we show that by virtue of the network geometry, a weak topological superconducting phase can develop, hosting spin-polarized weak topological boundary modes. 
This demonstrates that a ferromagnetic spin configuration of the decorated superconducting network can generate a TSC.
Unexpectedly, we also find that in some regions of the topological phase space, the topological boundary modes can dissociate from the bulk electronic bands, leading to a bulk-dissociated topological state. Such bulk-dissociated topological phases can be rigorously defined by their susceptibility to Anderson disorder e.g. when boundary modes are localizable by disorder and local boundary perturbations~\cite{Altland2024}.
By modulating the Fermi energy, we identify the emergence of nodal topological superconductivity where the bulk superconducting gap can be suppressed and, yet, topological edge bands remain intact.
We also find that low-energy corner modes can be found in the YSR network described by a hybrid weak TSC phase-- a weak TSC phase with both edge and corner states. We find these corner modes have properties such as an anomalously rapid convergence of energy scaling with system size which indicates the corner modes are also dissociated from higher-dimensional bands. Lastly, we show how manipulating boundary morphology can control the coupling between corner, edge, and bulk modes, highlighting how hetero-dimensional nanoscale metamaterial systems can serve as a useful template for controlling the coupling between electronic modes of different dimensionalities.

\section{Results}
\subsection{Yu-Shiba-Rusinov networks}
We model magnetic adatoms in a superconducting network as magnetic impurities with spin $S$ using an exchange interaction with coupling $J$ at the sites of the impurities. The system is defined on a real space Kondo lattice forming a square network with a lattice constant $a$ and plaquette constant (superlattice constant) $\ell$. Here we assume $S$ is sufficiently large relative to $J$ to treat the spin classically and ignore Kondo screening~\cite{Balatsky2006}. Magnetic impurities are assumed to be situated at the vertices of the network, see Fig.~\ref{fig:fig1}a. The superconducting state is described using the Bogoliubov de-Gennes (BdG) method with a mean-field superconducting gap $\Delta$. 
Additional details on the tight binding Hamiltonian are presented in Appendix A. 
While a bare square network does not possess topological phases since a square lattice by itself does not have Dirac points when spin is a good quantum number, in YSR networks, the decorated network superlattice leads to additional pseudospin degrees of freedom. With a sublattice structure, the superlattice generates an effective (pseudo)spin-orbit coupling from inter-sublattice coupling (see Appendix B). This effective spin-orbit coupling heuristically describes band inversion the network.

We can characterize the bulk topology by calculating the strong topological invariant. According to the Altland-Zirnbauer classification, YSR superlattices belong to class D and the Chern number is the corresponding topological invariant. Using the bulk band structure, we find the Chern number is equal to zero (trivial) for all the calculations we focus on in this work since we ignore any spin-orbit interactions or spin textures.
In Appendix D, we discuss the generation of non-zero Chern numbers in the presence of Rashba SOC.
Primarily, we investigate weak topological phases where the bulk topology is sensitive to boundary morphology. For two-dimensional superconductors in class D, we have $\mathbb{Z}_2$ weak topological symmetry indicators $v_{x,\pi}$ and $v_{y,\pi}$ constructed from Pfaffians evaluated at high-symmetry points in the Brillouin zone~\cite{Fu2007, Seroussi2014}:
\begin{align} 
    (-1)^{v_{x,\pi}} & = -s_{(\pi/\ell, 0)} s_{(\pi/\ell, \pi/\ell)}, \\
    (-1)^{v_{y,\pi}} & = -s_{(0, \pi/\ell)} s_{(\pi/\ell, \pi/\ell)}, \\
    s_{{\bf K_i}} & = \mathrm{sgn} \left( \mathrm{Pf}\left[ \tilde{H}({\bf K}_i) \right] \right),
\end{align} 
where $\tilde{H} = U^{\dagger} H U$ is the YSR network Hamiltonian in a basis where $\tilde{H}$ is an antisymmetric matrix and $U$ is a unitary matrix.
When $v_{x,\pi}$ ($v_{y,\pi}$) is equal to one, x-edge (y-edge) hosts edge states; otherwise, the edge is trivial. 

In Fig.~\ref{fig:fig1}b, we present the normal state dispersion of the YSR network with impurity spacing $\ell=10 a$.
A mini-Brillouin zone is formed with boundaries $\pm \pi / \ell$ owing to the network superlattice.
Curiously, we observe an apparent particle-hole symmetry in the normal-state dispersion of the network (even for $JS=0$) when $\ell$ exceeds some threshold value (nominally $7a$); for $\ell \le 7a$, the apparent particle-hole symmetry of the dispersion is lifted.
The flat band at half filling emerges as a consequence of the superlattice constant $\ell$ ($\gg a$) of the network creating an interference of modes in the square plaquettes, leading to localized states. Similar flat band features have been reported before in metallic networks~\cite{Lee2020}. In our work, we will focus on less than half filling of the total bandwidth ($1.5t \le E_F \le 3t$). When we turn on superconductivity ($\Delta > 0$), YSR bound states tend to develop at magnetic impurity sites. An example of a YSR bound state in a 1D superconducting chain is shown in Fig.~\ref{fig:fig1}c, displaying short wavelength oscillations enveloped by an exponential suppression with characteristic length scale $\xi_{ysr}$. In the simulations we present in this work $\xi_{ysr}$ varies between $35a$ and $45a$, always exceeding the superlattice constant $\ell$.

In Fig.~\ref{fig:fig1}d-i, we present the evolution of weak topological indicators and spectral features of the YSR network as we sweep the Fermi energy $E_F$ with fixed $JS = 0.6t$ and $\ell = 10a$.
Here we characterize electronic phases of the network by calculating topological indicators and spectral gaps from the bulk band structure $\{E_n(k_x,k_y) \}$, edge dispersion $\{E_{n,m}(k_x) \}$ with open boundary conditions along the y-axis, and the bound state spectrum of a finite-sized network $\{E_{n,m,l} \}$ where $n,m,l$ are the usual principal quantum numbers.
We identify three distinct phases using energy scales illustrated in Fig.~\ref{fig:fig1}d: phase I where $v_{x/y,\pi}=+1$, $\Delta_b > 0$ and $\delta_{SEB}=0$; phase II where $v_{x/y,\pi}=+1$, $\Delta_b \ge 0$ and $\delta_{SEB}>0$; and phase III where $v_{x/y,\pi}=+1$, $\Delta_b > 0$ and $0 \le \delta_0 < \delta^*$. Note that $\delta_{SEB}$ is the direct spectral gap separating edge and bulk bands.
A summary of these weak TSC phases is given in Table~\ref{tab:phases}.
Phases with $\delta_{SEB}>0$ can be viewed as \textit{bulk-dissociated} in the sense that the usual edge-bulk band hybridization occurring at the bulk gap edge is absent.
In Fig.~\ref{fig:fig1}e, we present a calculation of $v_{x,\pi}$ and $v_{y,\pi}$ versus the Fermi energy $E_F$. We see that the YSR network is in the weak topological phase with isotropic weak topological indicators: $v_{x,\pi} = v_{y,\pi} \equiv v_{\pi}$.
In Fig.~\ref{fig:fig1}f, we present the evolution of the bulk gap with $E_F$ where $\Delta_b$ oscillates between zero and amplitudes that exceed the superconducting gap $\Delta$ used in the BdG Hamiltonian.
Comparing Fig.~\ref{fig:fig1}e-i, at $E_F = 1.5t$ we see the network is in phase I with $v_{\pi} = +1$ and a mass gap $\delta_0+\delta^*>0$ in the edge state spectrum. 
With increasing $E_F$, the system enters phase II with a detaching of the edge state from the bulk bands, $\delta_{SEB}>0$. The peak magnitude of $\delta_{SEB}$ coincides with the closure of the mass gap $\delta_0+\delta^*$. Around $E_F\approx 1.85t$, the bulk gap closes and $v_{\pi} = 0$ (edge modes vanish).
Note that changes in the value of $v_{\pi}$ always coincides with the bulk gap closing or opening.
Near $E_F \approx 1.9t$, the bulk gap re-opens and the system enters phase III with bound states below the edge mass gap and $\delta^*>0$. These bound states have their spectral density pinned to the corners of the YSR network.
Then at $E_F \approx 2.2t$, $\delta^*$ and $\delta_0$ close, and the network enters phase II$^*$ where $v_{\pi} = +1$ but $\Delta_b = 0$.
In phase II$^*$, the Chern number is no longer well-defined, but the weak topological indicators related to the Pfaffian of the Hamiltonian remain well-defined; although, topological indicators are not guaranteed to be reliable. Despite this, in the simulations we present in this work, we observe empirically that $v_{x,\pi}$ and $v_{y,\pi}$ are, indeed, reliable when the bulk gap is closed and $\delta_{SEB}>0$.

\begin{table}[t]
	\centering
	\begin{tabular}{||c || c | c | c | c | c ||} 
		\hline
		Weak topological superconductor & $\Delta_b$ & $\delta_{SEB}$ & $\delta^*$ & $\delta_0$  \\ [0.5ex] 
		\hline
		\hline
		First-order bulk-coupled (I) & $> 0$ & $0$ & $\ge 0$ & DNE \\
		\hline
		First-order bulk-dissociated (II) & $\ge 0$ & $> 0$ & $\ge 0$ & DNE  \\
		\hline
		Hybrid bulk-coupled (III) & $> 0$ & $0$ & $> 0$ & $0 \le \delta_0 < \delta^*$ \\
		\hline
		Hybrid bulk-dissociated (IV) & $\ge 0$ & $\ge 0$ & $> 0$ & $0 \le \delta_0 < \delta^*$ \\ 
		\hline
	\end{tabular}
	\caption{\label{tab:phases}
		The spectral classification of class D topological superconducting phases discussed in this work based on spectral gaps in the thermodynamic limit. DNE is an abbreviation for Does Not Exist, and hybrid weak TSCs are defined by the co-existence of topological edge modes and non-topological corner modes.
        Phase (II$^*$) has the characteristics of phase (II) but $\Delta_b = 0$, and phase (IV$^*$) has the characteristics of phase (IV) but $\delta_{SEB} = 0$.}
\end{table}

In Fig.~\ref{fig:fig1}j, we present the density of states (DOS) versus $E_F$ of the YSR network in a nanoribbon geometry with translational invariance along the x-direction and open boundary conditions in the y-direction.
This allows us to resolve the DOS of both bulk and edge states.
For $E_F \le 2t$, we observe out-of-phase oscillations of the bulk and edge gaps with gradual band broadening, leading to oscillating regions with a finite $\delta_{SEB}$ shown in Fig.~\ref{fig:fig1}g. For $E_F > 2t$, the bulk gap continues to modulate, but new bump features appear at low energies, indicated by (*), that coincide with the onset of phase II$^*$ where $v_{\pi} = +1$, $\Delta_b = 0$, and $\delta_{SEB}>0$. The peaks in DOS indicated by (*) are associated with slow moving excitations in the edge state band. Given the spectral characterization of these various bulk-dissociated phases, we will now turn to study how bound states in weak TSC and bulk-dissociated phases can be characterized and manipulated.

\begin{figure}[h!!!]
    \centering
    \includegraphics[width=0.99\linewidth]{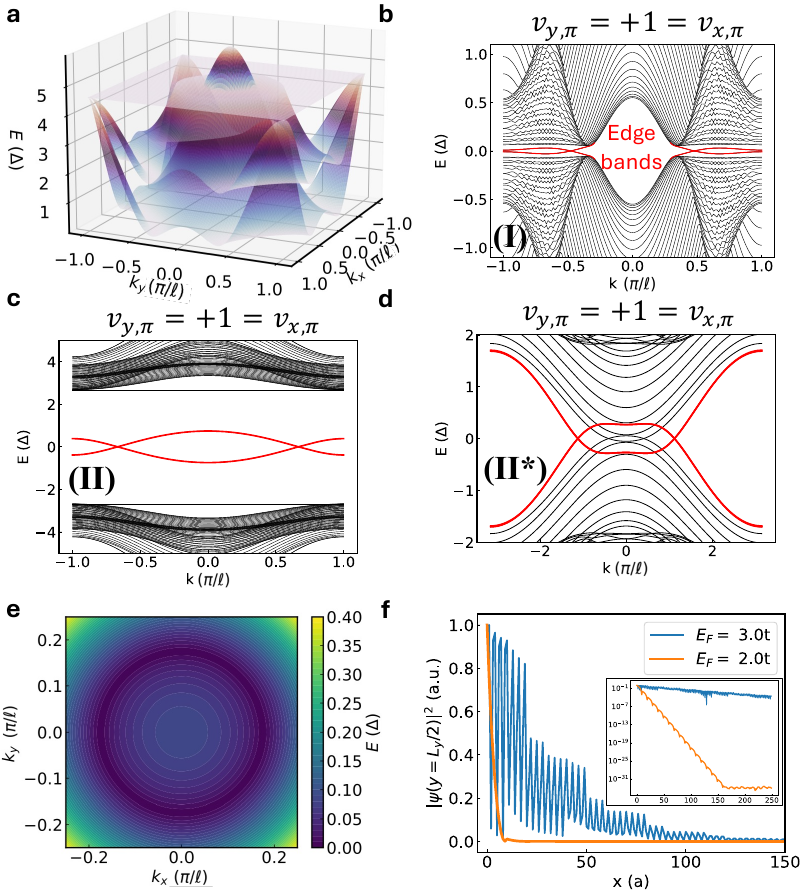} 
    \caption{\label{fig:fig2} 
    \textbf{a} The BdG band structure of the YSR network with $\ell = 10a$, $E_F = 3t$, and $JS = 0.8t$.
    \textbf{b} The YSR network dispersion with weak edge bands (red) connected to the bulk excitation spectrum using $\ell = 10a$, $JS=0.8t$ and $E_F = 3t$.
    \textbf{c} The YSR network dispersion with a separated edge band (red) using $\ell = 10a$, $JS=0.6t$ and $E_F = 2t$.
    \textbf{d} The YSR network dispersion with separated edge bands (red) dissociated from the nodal bulk excitation spectrum, and
    \textbf{e} the corresponding color map of the bulk gap for a YSR network with $\ell = 10 a$, $E_F = 2.22t$ and $JS = 0.6t$.
    \textbf{f} A cross-sectional cut of the local density of states of edge modes eigenstate in the bulk-coupled ($E_F = 3t$, $JS=0.8t$) and bulk-dissociated ($E_F = 2t$, $JS=0.6t$) phases.
    }
\end{figure}

\subsection{Bulk-dissociated topological edge states}
We present the excitation band structure in Fig.~\ref{fig:fig2}a using $\ell = 10a$, $E_F = 3t$ and $JS = 0.8t$, where we see a flat band emerge near $E=4\Delta$, characteristic of a network geometry~\cite{Lee2020}. 
Unless otherwise stated, we take $\Delta = 0.04 t$.
The $C_4$ rotational symmetry of the band structure reflects the $C_4$ symmetry of the square network.
In Fig.~\ref{fig:fig2}b, we present the dispersion of a YSR network nanoribbon with translational invariance (periodic in $\ell$) along the x-direction and open boundary conditions along the y-direction.
Calculating weak topological indicators, we find $v_{\pi} = +1$ which implies the gapless bands are, indeed, topological edge bands belonging to phase I where $\delta_{SEB} = 0$. 
We find the edge state dispersion is spin-polarized, two-fold degenerate, gapless and merges with the bulk bands.
In Fig.~\ref{fig:fig2}c, we show the dispersion of the YSR network nanoribbon with a lower Fermi energy, $E_F = 2t$, where the gapless edge bands are now separated from the bulk bands at all momenta in the Brillouin zone.
Note that the exchange term in the YSR network Hamiltonian breaks time-reversal symmetry so that crossings at time-reversal invariant momenta are no longer protected by Kramers' theorem.
This indicates the system is in phase II where $\delta_{SEB}>0$.
Next, we consider a YSR network with $E_F = 2.22t$.
Figure~\ref{fig:fig2}d shows gapless edge bands are dissociated from the bulk bands that cross the Fermi energy in phase II$^*$ where $\Delta_b = 0$. Here, bulk dissociation of the edge bands presents with the lack of edge-bulk hybridization at momenta where edge and bulk bands cross. The bulk energy gap near the $\Gamma$ point in the YSR network Brillouin zone is shown in Fig.~\ref{fig:fig2}e where we see that a nodal-line superconductor-- a superconductor with a suppressed superconducting gap along a continuous line in k-space-- develops with an isotropic suppression of the bulk superconducting gap. The resilience of the edge band with $\Delta_b = 0$ is a consequence of suppressed hybridization of bulk and edge bands, signaling the bulk gap needs not be finite with bulk-dissociated edge states. 
\begin{figure*}[tbp]
	\centering
	\includegraphics[width=0.95\linewidth]{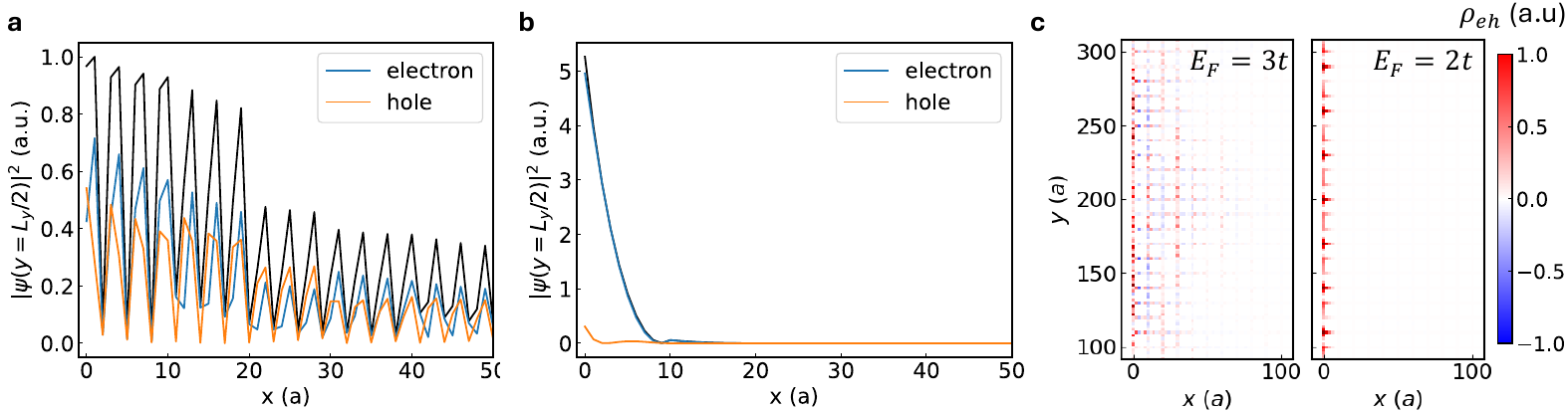} 
	\caption{\label{fig.r07} 
		\textbf{a} A cross-sectional cut of a edge modes eigenstate (black) in the bulk-coupled ($E_F = 3t$, $JS=0.8t$) phase with its electron (blue) and hole (orange) amplitudes, where $\vert \psi_e({\bf r})\vert^2 = \psi^{\dagger}({\bf r}) \left(\frac{\tau_z+\tau_0}{2} \right) \psi({\bf r})$ and $\vert \psi_h({\bf r})\vert^2 = \psi^{\dagger}({\bf r}) \left(\frac{\tau_0-\tau_z}{2} \right) \psi({\bf r})$, for Pauli matrices $\tau_i$ in particle-hole space (see Appendix A for more information).
        \textbf{b} A cross-sectional cut of a edge modes eigenstate (black) in the bulk-dissociated ($E_F = 2t$, $JS=0.6t$) phase with its electron and hole amplitudes.
		\textbf{c} Colormaps for $\rho_{eh}({\bf r})$ near the edge of the two systems.
	}
\end{figure*}

To distinguish bulk-coupled and bulk-dissociated edge modes, we calculate the local density of states (LDOS) of the lowest positive energy eigenstate of a finite-sized YSR network. Figure~\ref{fig:fig2}f shows a cross-sectional cut of LDOS of bulk-coupled and bulk-dissociated edge eigenmodes. The conventional weak TSC edge eigenstate with $E_F = 3t$ and $JS=0.8t$ is confined to the edge of the network with a bulk penetration depth of $\sim 5\ell$. On the other hand, the bulk-dissociated edge mode with $E_F=2.22t$ and $JS=0.6t$ (a similar result is found with $E_F = 2t$) has a large spectral weight at the boundary with bulk penetration $\sim \ell$ before a complete suppression of the LDOS further into the bulk, see inset of Fig.~\ref{fig:fig2}f. This can be interpreted as a consequence of the lack of spectral flow between bulk valence and conduction bands via the edge in the following way. For a Chern insulator forming a Corbino disk with Chern number $C$, the adiabatic application of an external magnetic flux $\Phi$ threading the disk will pump a quantized charge $Ce$ along the radial direction of the disk when $\Phi$ reaches an integer multiple of $h/e$, where $e$ is the fundamental electric charge and $h$ is Planck's constant~\cite{Laughlin1981}. This phenomenon is a consequence of spectral flow and ensures the edge band remains gapless. In the case of a bulk-dissociated edge band, heuristically two energy gaps can prevent this: $\delta^*$ and $\delta_{SEB}$ (see Fig.~\ref{fig:fig1}d). Then in phase II, we postulate $\delta_{SEB}>0$ acts as an energy barrier between edge modes and bulk electronic degrees of freedom, completely eliminating the LDOS of edge modes deep in the bulk as shown in the inset of Fig.~\ref{fig:fig2}f. 
We verified in Appendix C that weak structural disorder does not eliminate bulk-dissociated edge modes with a finite bulk gap, $\Delta_b > 0$.
Note that, realistically, disorder will inevitably lead to coupling of edge and bulk states at band crossings in phase II$^*$ due to electronic scattering, but phase II$^*$ is still unique in that flux pumping of quantized charge between edges via spectral flow is nevertheless prohibited. This illustrates the novelty of bulk-dissociated weak TSCs since such spectral flow in strong TSCs in two dimensions is guaranteed~\cite{Altland2024, Nakamura2025_prl}.
As we will explore in the next section, this will impact the scaling properties of bulk-dissociated boundary modes and lead to a clear distinction between bulk-coupled and bulk-dissociated topological phases.

Lastly, we analyze the weight of electron and hole components of bulk-dissociated edge eigenstates. In the YSR network, the coupling of sublattice pseudospins hybridizes and splits electron bands (and separately, hole bands), which promotes interband pairing between sublattice psuedospins. For strong interband pseudospin hybridization, electron and hole bands can be pushed to zero energy in the superconducting state. To get a sense of how this type of hybridization affects our results, we look at the particle and hole content of edge modes in the weak topological phases of our YSR networks with a gapless edge dispersion.
The particle and hole content of edge modes in Fig. 2f of the manuscript is shown in Fig.~{\ref{fig.r07}}. We see that for the weak edge eigenstate in the bulk-coupled phase in Fig.~{\ref{fig.r07}}a, we see a nearly equal balance of electron and hole components with a slightly larger weight toward electron character. For the bulk-dissociated edge eigenstate in Fig.~{\ref{fig.r07}}b, the edge mode has a much larger weight in the electron component. In Fig.~{\ref{fig.r07}}c, we present the real-space colormap of the difference in electron and hole components,
$
\rho_{eh}({\bf r}) = \Psi^{\dagger}({\bf r}) \tau_z \Psi({\bf r}),
$
in the two cases. The significant electron-hole imbalance in bulk-dissociated edge modes suggests interband pseudospin superconducting pairing plays an important role in the bulk-dissociated weak topological phase.

\subsection{Bulk-dissociated corner modes}
\begin{figure*}[tbp]
	\centering
	\includegraphics[width=0.95\linewidth]{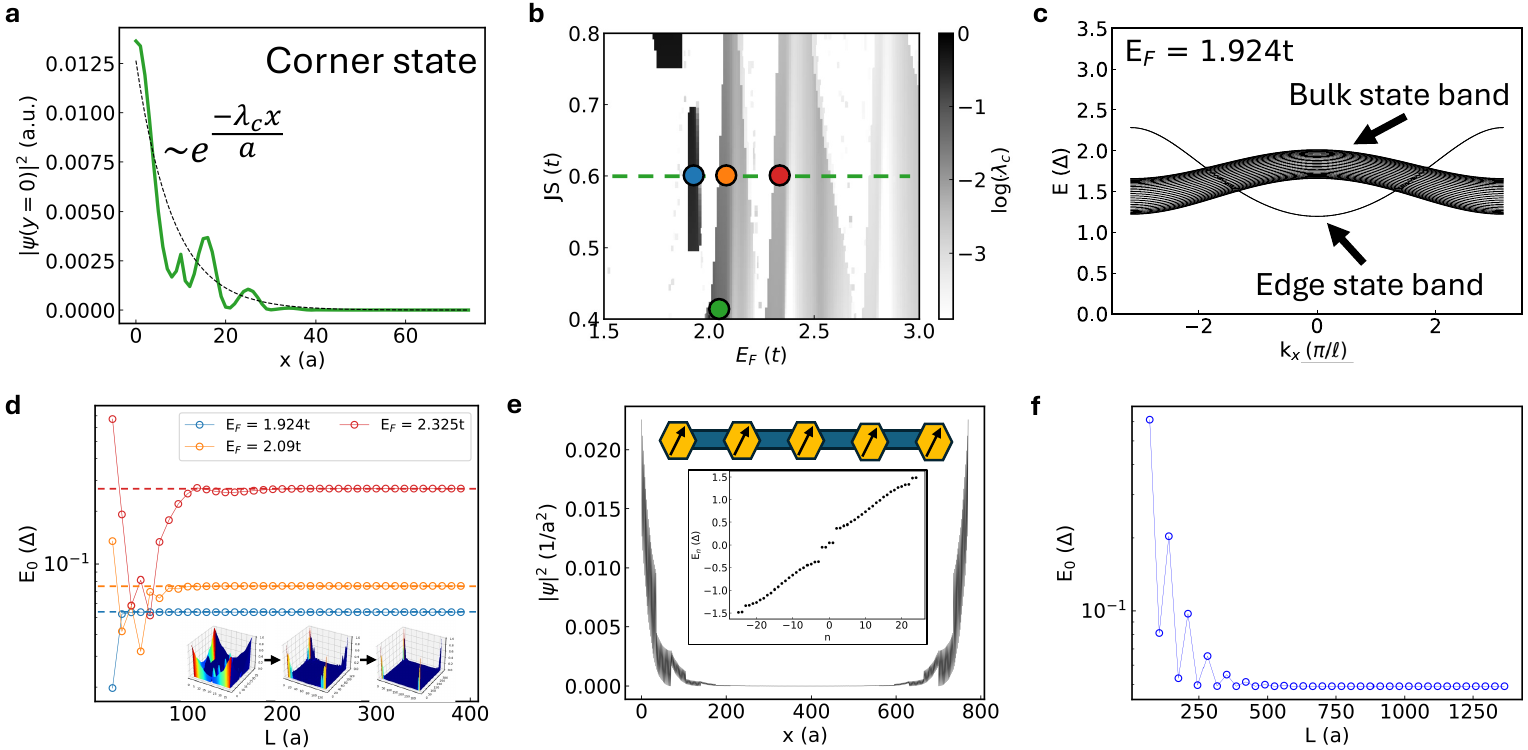} 
	\caption{\label{fig:fig3a}
		\textbf{a} The local density of states of a corner mode along $y=0$ using $JS=0.4 t$ and $E_F = 2.1 t$. The dashed line is an exponential fit of the wavefunction with characteristic scale $\lambda_c$.
        \textbf{b} The corner mode phase diagram for $\log(\lambda_c)$ versus $JS$ and $E_F$.
		\textbf{c} The YSR network dispersion with weak edge bands crossing the bulk excitation band using $\ell = 10a$, $JS=0.6t$ and $E_F = 1.924t$.
		\textbf{d} The lowest positive bound state energy versus $L$ for an $L\times L$ YSR network using $\ell = 10a$, $JS=0.6t$ and $E_F = 1.924t,~ 2.09t,~ 2.325t$. Inset: Local density of states of corner mode with progressively larger $L$.
		\textbf{e} The local density of states of delocalized Andreev bound state in the YSR chain ($L_y = 1$) with $\ell = 35a$, $JS=0.8t$ and $E_F = 2t$. Inset: Eigenspectrum $E_n$ of modes in the chain.
		\textbf{f} The delocalized Andreev bound state energy in the YSR chain versus $L$.
	}
\end{figure*}

Next we will investigate the appearance of bound states below the edge band mass gap $\delta^*$ in phase IV$^*$ in Table~\ref{tab:phases} where boundary modes are bulk-dissociated but $\delta_{SEB}=0$. The bound states in this phase are found to be highly localized corner modes in the network
and below we argue that these are not Majorana corner modes.
In Fig.~\ref{fig:fig3a}a, we present a line cut at $y=0$ of the LDOS of a corner mode using $JS=0.4t$ and $E_F = 2.1t$. We again emphasize that no spin-orbit interactions are needed to realize this exotic phase. The positive energy corner modes are four-fold degenerate and spin-polarized, indicating they are not spinless Majorana modes. The four-fold degeneracy here arises from the $C_4$ symmetry of the square network. To check whether the corner modes are gapped out by a modification to the boundary, we consider the effects of attaching a trivial superconducting region with depth $L_{boundary}$ to the bottom of the network. Coupling the uniform superconducting region to the base of the square network eliminates possible quantum interference effects arising from the network that may pin localized modes at the corners by lifting the $\mathcal{M}_y$ mirror symmetry of the network. Furthermore, we expect topological boundary modes forming at the interface with a vacuum to be robust to weak coupling with a trivially gapped system~\cite{Peterson2018}. The inset of Fig.~\ref{fig:fig3a}a shows that, despite direct coupling to the trivial superconductor, the bottom corner modes indeed remain pinned to the base of the network and the top corner modes are largely unaffected. First, this suggests the corner modes do not arise from a trivial geometric effect of the network, but rather, are protected by the edge mass gap present in the YSR network. 
Second, the distinct behavior of top (unperturbed) corner modes from bottom (hybridized) corner modes can be viewed as a signature of non-topological corner modes. In the case of Majorana corner modes, while each corner can occupy a Majorana fermion quasiparticle, this fermionic system must have a fermion occupation number that is conserved. As a consequence, Majorana bound states must be non-local states occupying an even number of corners. In contrast, the corner modes we study here can be manipulated to occupy a single corner by adding a trivial superconductor which only interfaces with one corner (see Fig.~\ref{extended:fig1}).

We characterize the localization of the corner mode with an exponential fit $\sim e^{-\lambda_c x/a}$ and extract the characteristic scaling factor $a/\lambda_c$. 
Here, $\lambda_c$ is a unitless quantity with two limits: $\lambda_c \to 0$ describing a delocalized state, and $\lambda_c \to +\infty$ describing a state localized on a single site at the corner. Now, in our simulations, $a/\lambda_c$ is usually on the order of $10a$ implying $\lambda_c$ is on the order of $0.1$, whereas strong confinement of the spectral weight corresponds to $\lambda_c = 1$.
We can use $\lambda_c$ to characterize corner states, similar to using Lyapunov exponents to characterize Majorana zero modes~\cite{DeGottardi_2011, DeGottardi2013,DeGottardi2013_prl, Hegde2016}. By calculating $\log(\lambda_c)$ of the lowest positive energy bound states of the YSR network, we generate a phase diagram for corner modes in Fig.~\ref{fig:fig3a}b. We observe large regions of phase space hosting corner modes with a varying degree of localization characterized by $\lambda_c$. For $\log(\lambda_c) \lesssim 0$ when $E_F < 2t$, the penetration of the corner state into the bulk is less than $\ell$, suggesting bulk dissociation. For $E_F>2t$, the corner modes have a less concentrated spectral weight that roughly continues to decrease with increasing $E_F$. This implies that highly localized spectral weight alone cannot necessarily distinguish these corner modes from Majorana corner states.

In Fig.~\ref{fig:fig3a}c we present the dispersion of a YSR network nanoribbon when $E_F = 1.924t$ and $JS = 0.6 t$. Here we see an edge band is present, consistent with $v_{\pi} = +1$, and it crosses a band of bulk states. Here $\delta_{SEB} = 0$ and edge and bulk bands cross, which leads to mixing of the edge and bulk bands when disorder is present. In Fig.~\ref{fig:fig2}, we showed bulk-coupled and bulk-dissociated TSC phases can be distinguished by the spectral weight concentration of edge modes. Now we show that this also holds for corner modes with a distinctive bound state energy scaling with system dimension $L$. Figure~\ref{fig:fig3a}d shows the scaling of the energy of a corner mode $E_0$ with $L$ for $JS=0.6 t$ and $E_F = 1.924t,~ 2.09t,~ 2.325t$, where we observe a rapid convergence of the bound state energy leading to an anomalous $L$-independent energy for $L > 200a$. This differs drastically from the expected exponential scaling of Majorana corner modes~\cite{DasSarma2012} and power law or exponential scaling of corner modes in gapless higher-order topological insulators~\cite{Verresen2021}. We also note the intuitive correlation between the length where $E_0$ converges, $L_c$, and $\lambda_c$: smaller $\lambda_c$ leads to larger $L_c$. In this work, we will use this asymptotic $L$-independent energy scaling to characterize bulk-dissociated corner modes.

We can also analyze corner modes in a simpler limit where we collapse one of the dimensions of the network creating a dilute spin-polarized 1D YSR chain without spin-orbit coupling. In Fig.~\ref{fig:fig3a}e, we show the LDOS of a non-topological boundary mode in the YSR chain. In this case, the two-fold degenerate boundary modes are spin-polarized and the strong topological invariant (the Pfaffian in 1D) is trivial. The two-fold degeneracy is protected by the mirror symmetry of the chain. These modes were first observed by K\"{u}ster \textit{et al.} where no spin-orbit coupling was needed to generate non-Majorana end modes in a dilute spin chain. Similarly, the interference of weakly coupled YSR states with Rashba spin-orbit coupling has been shown to also result in non-topological boundary modes in YSR chains that can mimic signatures of MZMs~\cite{Kuster2022, Hess2023}. 
This highlights how YSR networks in the weak TSC phase are not correlated with the presence of a strong topological phase in their constituent 1D chains.
In Fig.~\ref{fig:fig3a}f, we show the scaling of the boundary mode energy with the YSR chain length $L$. We find the boundary mode converges rapidly with $L$, similar to the corner states in the YSR network shown in Fig.~\ref{fig:fig3a}d. This suggests the non-topological boundary modes in the dilute YSR chain have the same origin as bulk-dissociated corner modes in the YSR network.

\begin{figure}[tbp]
	\centering
	\includegraphics[width=0.99\linewidth]{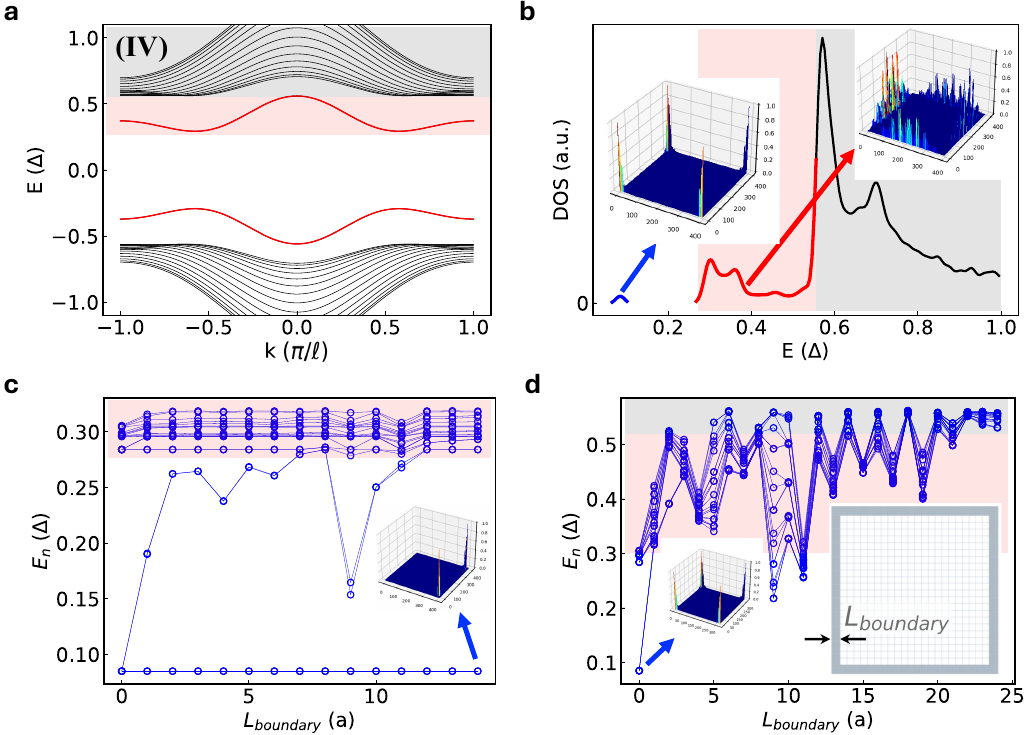} 
	\caption{\label{fig:fig3} 
		\textbf{a} The YSR network dispersion with a separated edge band using $\ell = 18a$, $JS=0.8t$ and $E_F = 2.2t$.
		\textbf{b} The density of states versus $E$ showing both separated edge states and corner modes simultaneously exist.
		\textbf{c} Eigenenergies $E_n$ versus the thickness $L_{boundary}$ of a trivial superconductor boundary attached to the bottom of the YSR network.
		\textbf{d} Eigenenergies $E_n$ versus the thickness $L_{boundary}$ of a trivial superconductor boundary attached to the perimeter of the YSR network.
	}
\end{figure}

To further explore corner modes in this system, we increase $\ell$ and identify a hybrid bulk-dissociated TSC phase with $\delta_{SEB}>0$ and $\delta_0 + \delta^* > 0$ in Fig.~\ref{fig:fig3}a-b.
Figure~\ref{fig:fig3}a shows the edge dispersion using $\ell = 18a$, $JS=0.8t$ and $E_F = 2.2t$ where the edge band is dissociated from the bulk bands and the edge bands also have a finite mass gap. Here $v_{\pi} = +1$ indicates a weak TSC. The density of states of an $L$x$L$ YSR network with $L=400a$ is shown in Fig.~\ref{fig:fig3}b. We find localized states lie below the edge band gap near zero energy which are highly confined to the corners of the network, as shown in the inset of Fig.~\ref{fig:fig3}b.
Since $\delta_{SEB} > 0$, this belongs to phase IV in Table~\ref{tab:phases}.
Similar to the approach described for the inset of Fig.~\ref{fig:fig3a}a, we can characterize the corner modes' robustness to deformations of the boundary by attaching a trivial superconductor to the bottom of the network.
In Fig.~\ref{fig:fig3}c, we present the positive bound state energies versus $L_{boundary}$ and notice that the lower corner modes coupled to the trivial superconductor are gapped out for $L_{boundary} > 10a$ while leaving the upper pair of corner modes unaffected. The bottom corner modes dissolve into edge modes confined to the bottom of the network when $L_{boundary} = 11a$ unlike the corner modes analyzed in the inset of Fig.~\ref{fig:fig3a}a. This is due to the small gaps $\delta^*$ and $\delta_{SEB}$ in Fig.~\ref{fig:fig3}b protecting the hybridization of modes of different dimensionality. This illustrates how boundary morphology of the network can be manipulated to hybridize corner and edge modes. This idea can be extended further to couple boundary modes to bulk, as shown in Fig.~\ref{fig:fig3}d. Here, we attached a trivial superconducting region to the entire perimeter of the network and observe both corner and edge modes are pushed into the bulk band near $0.5 \Delta$ and hybridize. Hence, manipulating boundary morphology of a YSR network can help distinguish topological and trivial modes while also being a useful tool to control the hybridization of bulk and boundary modes.

\section{Discussion}
Edge bands disconnected from the bulk have been noted in weak topological superconducting phases in two-dimensional Shiba lattices with Rashba spin-orbit interactions and stacked magnetic structure~\cite{Wong2023}, and topological insulators with altermagnetism~\cite{Li2025}. 
Recent studies have established a general classification revealing that most three-dimensional strong topological insulators and superconductors—excluding Wigner-Dyson classes A, AI, and AII—are susceptible to surface-bulk dissociation and localization~\cite{Altland2024, Nakamura2025_prl, Shiozaki2025_prb, Lapierre2026}. A variety of one-dimensional topological phases were also shown to be localizable; meanwhile, all the two-dimensional topological phases in the Wigner-Dyson classification were found to be non-localizable. Here, we have shown that this classification does not necessarily translate to weak topological phases.
While the system in Ref.~{\onlinecite{Li2025}} reported separated edge bands in a strong topological phase, it required stringent fine-tuning to generate bulk-dissociated phases with systems with spin-orbit interactions. Here we report bulk-dissociation occurring in a \textit{weak} topological superconducting phase \textit{without} any intrinsic or extrinsic spin-orbit coupling. With our fundamentally different approach, we identify a novel gapless weak topological phase, and our system generally defies the usual heuristic argument for weak phase construction since constituent 1D chains forming the network are topologically trivial when isolated from the network.

We also find that bulk-dissociated topological bands represent more than a simple detachment from the bulk bands with $\delta_{SEB}>0$; they can lead to unexpected nodal topological superconducting phases and unusual scaling properties of topological bound states.
Moreover, none of the topological phases identified this work required any SOC terms in the Hamiltonian. Creating TSCs hosting Majorana zero modes will necessitate SOC, but the network geometry may be useful to create better protected Majoranas when the SOC is weak.
In Appendix D, we explicitly show that YSR networks with Rashba spin-orbit coupling can host chiral Majorana zero modes with a non-zero Chern number.
This could be helpful to address the issue of weak SOC in superconducting hybrid heterostructures discussed in the Introduction.
On the other hand, we show in pristine systems with a superlattice structure how boundary modes that mimic Majorana modes can be generated. To our knowledge, this is the first construction of trivial corner modes that mimic higher-order topological Majorana corner modes. Our characterization of these topologically-trivial bulk-dissociated modes should help rule out false-positives of genuine Majorana modes.

Our results also highlight the subtle role of bulk-boundary coupling in topological systems. The YSR network we studied in this work can be viewed as a \textit{hetero-dimensional nanoscale metamaterial} (HNM), where 1D components are geometrically assembled to form a 2D structure which is then decorated with 0D magnetic adatoms. We found this hetero-dimensional system serves as a useful template for controlling the coupling between modes of different dimensionalities i.e. coupling of 2D bulk, 1D edge, and 0D corner modes. Recently, there is a growing interest in studying this type of coupling which can give rise to novel phenomena in topological semimetal systems~\cite{Cuozzo_leggett, Zhang2024_surface_bulk, Yu2024, Guo2025} and analyzing the evolution of topological phases as system dimensionality is reduced in YSR lattices~\cite{Rodriguez2025}. Here we establish the YSR network as a useful system to study bulk-boundary coupling directly by manipulating boundary morphology.

\section{Conclusions}
In summary, we have shown that neither spin-orbit interactions nor spin textures are necessary to generate a wide variety of topological superconducting phases in magnetic adatom decorated superconducting networks.
We focused on a distinct flavor of topological superconductivity in which the bulk and boundary modes are dissociated spectrally. We found that a variety of interesting features can be realized in YSR networks in bulk-dissociated phases, such as a nodal-line weak topological superconducting phase and corner modes.
More broadly, our work highlights the nuanced role of bulk and boundary mode coupling in determining spectral properties of topological phases that may affect topological phases in other symmetry classes and dimensions.

In order to experimentally realize a YSR network, a bottom-up nanoassembly approach can be pursued. For instance, a HNM can be constructed by stacking a network of 1D metallic carbon nanotubes onto a 2D or 3D superconducting substrate, and decorating the network with magnetic impurities. The carbon nanotube network can be functionalized by attaching ligand shells encapsulating single-molecule magnets to the nanotube outer wall~\cite{Kyatskaya2009, Bogani2009, Urdampilleta2011}.
Alternatively, one can take advantage of superconducting networks emerging in twisted van der Waals materials. 
Here, the moire potential naturally creates an emergent conductive honeycomb network in the graphene bilayer.
In Appendix E, we verified that boundary modes indeed persist in a honeycomb YSR network without spin-orbit interactions.
Strong electronic correlations in magic-angle twisted bilayer graphene creates a 2D superconducting material on which single-molecules can be functionalized on the surface.  Similar approaches have been proposed before with non-magnetic adatoms~\cite{Skurativska2021}. In the context of YSR states, modulating the electron density in MATBG would allow one to control the strength of e-e correlations and tune between a strongly correlated Kondo phase and a weakly correlated phase where the magnetic impurities can be described classically.
Past work on topological properties of metallic networks suggest a honeycomb network may host even more robust topological properties than a square network owing to the additional sub-lattice degree of freedom~\cite{Lee2020}. Both approaches represent a more general construction of hybrid materials using a hetero-dimensional design. 
Moreover, the additional complexity in assembling hetero-dimensional nanostructures may be compensated by the advantage of a metamaterials design where the geometric engineering of the lower-dimensional materials can result in exotic phases with fewer intrinsic material property requirements (e.g. spin-orbit interactions). 
By capitalizing on advances in nanomaterial fabrication over the past decade, new approaches to low dimensional material nanoassembly may open a new door to a wide variety of new hybrid materials and increase accessibility to exotic topological phases of matter.
\\

\section*{Acknowledgments} 
J.J.C. thanks A. Cerjan for helpful discussions.
The work at Sandia is supported by a LDRD project. 
S.A.A.G. was supported by the U.S. Department of Energy, Office of Science, Basic Energy Sciences, under Award No. DE-SC0024942.
Sandia National Laboratories is a multi-mission laboratory managed and operated by National Technology \& Engineering Solutions of Sandia, LLC (NTESS), a wholly owned subsidiary of Honeywell International Inc., for the U.S. Department of Energy’s National Nuclear Security Administration (DOE/NNSA) under contract DE-NA0003525. This written work is authored by an employee of NTESS. The employee, not NTESS, owns the right, title and interest in and to the written work and is responsible for its contents. Any subjective views or opinions that might be expressed in the written work do not necessarily represent the views of the U.S. Government. The publisher acknowledges that the U.S. Government retains a non-exclusive, paid-up, irrevocable, world-wide license to publish or reproduce the published form of this written work or allow others to do so, for U.S. Government purposes. The DOE will provide public access to results of federally sponsored research in accordance with the DOE Public Access Plan.

\section*{Appendix}
\appendix

\begin{figure}[tbp]
	\centering
	\includegraphics[width=0.95\linewidth]{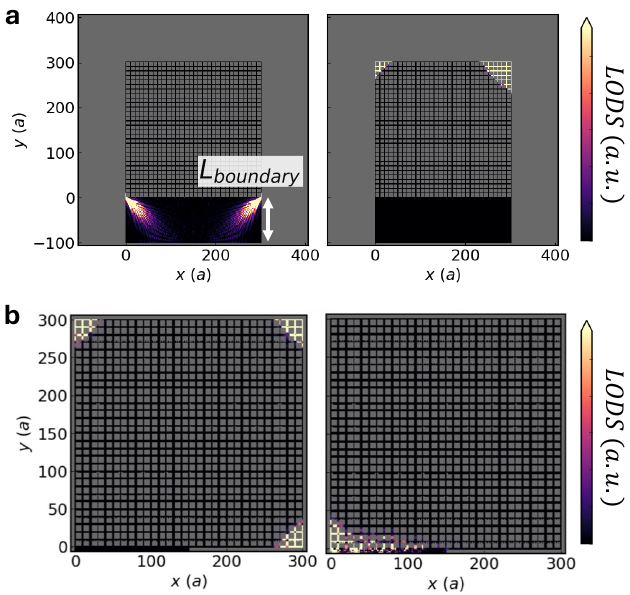} 
	\caption{\label{extended:fig1}
		Using the same parameters as were used to generate Fig.~\ref{fig:fig3a}a: (a) LDOS of corner modes with a trivial superconductor attached to the bottom of the network;
		(b) LDOS of corner modes with a trivial superconductor attached to the bottom left side of the network with $L_{boundary} = 20 a$.
	}
\end{figure}

\begin{figure}[tbp]
	\centering
	\includegraphics[width=0.7\linewidth]{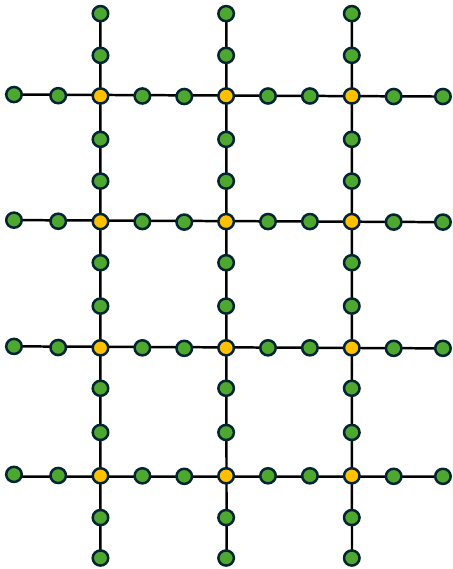} 
	\caption{\label{fig.r05} 
    Green sites denote superconducting sites with $JS = 0$ and yellow sites with $JS>0$.
	}
\end{figure}
\section{Methods}
We start by considering a one-dimensional metallic chain decorated with magnetic impurities. Let $\psi^{\dagger}({\bf r}) = \left( c_{\uparrow}^{\dagger}({\bf r}), c_{\downarrow}^{\dagger}({\bf r}) \right)$ be the fermionic spinor at site $\bf r$. The Hamiltonian on a Kondo lattice is
\begin{align}
    H & = \sum_{n=0}^N \psi^{\dagger}(n) \left( (2t - E_F) \sigma_0\right) \psi(n) \nonumber \\
    & + \sum_{n=0}^{N/\ell} \psi^{\dagger}(\ell n) \left( JS \sigma_z \right) \psi(\ell n) + \sum_{<i,j>} \psi^{\dagger}(i) \left( -t \sigma_0 \right) \psi(j),
\end{align}
where $t$ is the hopping, $E_F$ is the Fermi energy, $JS$ is the exchange interaction at the magnetic impurity sites, $\sigma_i$ are Pauli matrices in spin space, $N$ is the total number of sites in the chain, and $\ell$ is the magnetic impurity spacing. Throughout this work we assume $N/\ell \in \mathbb{N}$ and $\Delta = 0.04t$. Here we've taken the lattice constant $a=1$.

To model a YSR chain, we treat the superconductivity in the metallic chain within the Bogoliubov de-Gennes (BdG) formalism with a mean-field approximation of the superconducting gap $\Delta$. Here the BdG Hamiltonian satisfies the BdG equation:
\begin{align}
    H_{BdG} \Phi({\bf r}) = E \Phi({\bf r}),
\end{align}
where $\Phi$ is a wave function describing electron-like and hole-like excitations in the superconducting state, and $E$ is the excitation energy.
We assume $\Delta$ is spatially uniform and pick a gauge such that $\Delta > 0$. We take the spinor in Nambu space to be $\Psi^{\dagger}({\bf r}) = \left( c_{\uparrow}^{\dagger}({\bf r}), c_{\downarrow}^{\dagger}({\bf r}), c_{\uparrow}({\bf r}), c_{\downarrow}({\bf r}) \right)$. Then the BdG Hamiltonian in real space defining the YSR chain is
\begin{align}
    H_{BdG}^{(chain)} & = \frac{1}{2}\sum_{n=0}^N \Psi^{\dagger}(n) \left[ (2t - E_F) \tau_z \sigma_0 - \Delta \tau_y \sigma_y\right] \Psi(n) \nonumber \\
        & + \frac{1}{2} \sum_{n=0}^{N/\ell} \Psi^{\dagger}(\ell n) \left( JS \tau_z \sigma_z \right) \Psi(\ell n) \nonumber \\
        & - \frac{1}{2} \sum_{<i,j>} \Psi^{\dagger}(i) \left( t \tau_z \sigma_0 \right) \Psi(j) ,
\end{align}
where $\tau_i$ are Pauli matrices in Nambu space.
Extending the YSR chain model to a square network, we have
\begin{widetext}
\begin{align}
    H_{BdG}^{(network)} & = \frac{1}{2}\sum_{n_x=0}^N \sum_{n_y=0}^{N} \sum_{m=0}^{N/\ell} \Psi^{\dagger}(n_x, n_y) \left[ (4t - E_F) \tau_z \sigma_0 - \Delta \tau_y \sigma_y\right] \delta_{n_y,\ell m} \Psi(n_x, n_y) \nonumber \\
    & + \frac{1}{2}\sum_{n_x=0}^N \sum_{n_y=0}^{N} \sum_{m=0}^{N/\ell} \Psi^{\dagger}(n_x, n_y) \left[ (4t - E_F) \tau_z \sigma_0 - \Delta \tau_y \sigma_y\right] \delta_{n_x,\ell m} \Psi(n_x, n_y) \nonumber \\
    & + \frac{1}{2} \sum_{n_x=0}^N \sum_{n_y=0}^{N} \sum_{n,m=0}^{N/\ell} \Psi^{\dagger}( n_x, n_y) \left[ JS \tau_z \sigma_z - (4t - E_F) \tau_z \sigma_0 + \Delta \tau_y \sigma_y \right]\delta_{n_x,\ell n}~\delta_{n_y,\ell m} \Psi(n_x,  n_y) \nonumber \\
    &+ \frac{1}{2} \sum_{n_x=0}^N \sum_{n_y=0}^{N} \sum_{m=0}^{N/\ell} \Psi^{\dagger}( n_x +1, n_y) \left( -t \tau_z \sigma_0 \right) \delta_{n_y,\ell m} \Psi(n_x,  n_y) \nonumber \\
    &+ \frac{1}{2} \sum_{n_x=0}^N \sum_{n_y=0}^{N} \sum_{m=0}^{N/\ell} \Psi^{\dagger}( n_x, n_y+1) \left( -t \tau_z \sigma_0 \right)\delta_{n_x,\ell m} \Psi(n_x,  n_y)  + H.c.
    \label{eq:BdGHam}
\end{align}
\end{widetext}
Note that the last two terms in the square brackets of line three in Eq.~({\ref{eq:BdGHam}}) compensate for the double counting of the onsite terms arising from the preceding two lines.
A schematic of the lattice for networks is shown in Fig.~{\ref{fig.r05}}.
Numerical results were found using the Python package Kwant~\cite{Groth2014}. 
The unitary matrix used to transform Hamiltonian into antisymmetric matrix for Pfaffian calculations is $U = \frac{1}{\sqrt{2}} \begin{pmatrix} 1 & -i \\ 1 & i \end{pmatrix} \otimes \sigma_0$.
Unless otherwise stated, all calculations using open boundary conditions assume the edge termination occurs with no ``tails" so that the sites at the edge form a continuous chain. In band structure calculations and $\Delta_b$ calculations, the k-resolution is $0.02 \pi/\ell$ and calculations evaluating the scaling of the k-mesh have been performed to  validate the accuracy of gapless phases where physically $\Delta_b$ but numerically the gap remains small and finite due to the resolution of the k-sampling. We numerically define a gapless phase by $\Delta_b<0.01\Delta$. For edge dispersion calculations, the width is typically $400a$ and bound state spectrum calculations with open boundary conditions along x and y typically have system sizes ranging between $600a$ x $600a$ to $1000a$ x $1000a$; the system size required to eliminate finite size effects varies depending on $\ell$ and whether the system in a bulk-coupled or bulk-dissociated phase.
%

\begin{figure}[tbp]
	\centering
	\includegraphics[width=0.6\linewidth]{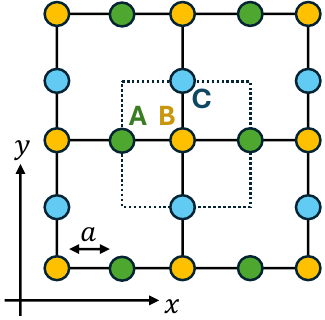} 
	\caption{\label{fig.r20} Cartoon of a YSR network with $\ell = 2a$ having a unit cell with sublattices A, B and C.
	}
\end{figure}

\section{Network sublattice analysis}
To analyze the generation of pseudospin degree of freedom in YSR networks, we consider the network in Fig.~{\ref{fig.r20}} where we label sublattice sites A, B and C. The presence of a multipartite lattice in YSR networks affords the system additional pseudospin degrees of freedom. To see this, we can write the Hamiltonian of the network in Fig.~{\ref{fig.r20}} in the following general form:
\begin{align}
    H_{BdG} &= \sum_{A_i,C_i} \left( \Psi_{A_i}^{\dagger} h_0\Psi_{A_i} + \Psi_{C_i}^{\dagger} h_0 \Psi_{C_i} \right)  \nonumber \\
    & + \sum_{B_i} \Psi_{B_i}^{\dagger} h_{JS} \Psi_{B_i}  +\sum_{\langle A_i,B_j \rangle} \Psi_{A_i}^{\dagger} h_{hop} \Psi_{B_j} \nonumber \\
    & +\sum_{\langle C_i,B_j \rangle} \Psi_{B_i}^{\dagger} h_{hop} \Psi_{C_j} + h.c.
\end{align}
Here $\Psi_{S_i}^{\dagger}$ and $\Psi_{S_i}$ are the creation and annihilation BdG spinor operators for the $i^{th}$ site on sublattice $S$, $h_0$ is the onsite matrix with $JS=0$, $h_{JS}$ is the onsite matrix with $JS>0$ describing the magnetic impurity, and $h_{hop}$ is the hopping matrix. As expected, hoppings only occur between sites on different sublattices. Furthermore, treating sublattice index as a pseudospin degree of freedom, we see the network generates an effective (pseudo)spin-orbit coupling from inter-sublattice coupling owing to the hopping terms. This spin-orbit coupling is anisotropic in the sense that the vectors connecting neighboring AB sites ($\{(\pm a,0)\}$) and BC sites ($\{(0,\pm a)\}$) are orthogonal.
%

\begin{figure}[tbp]
	\centering
	\includegraphics[width=0.99\linewidth]{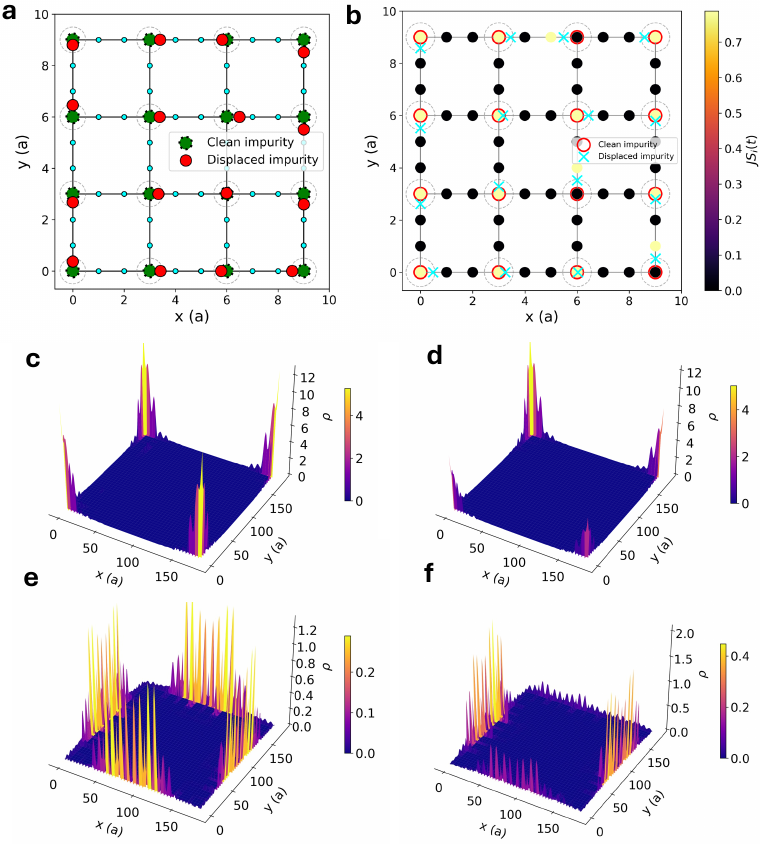} 
	\caption{\label{fig.r09} $JS$ is applied to the single site nearest to the displaced impurity position. $JS$ profile is a Dirac delta function. \textbf{a} A schematic for the disorder model and \textbf{b} schematic for the disorder $JS$ profile. \textbf{c} The lowest positive energy eigenstate confined to the corners in the pristine system. \textbf{d} Corner mode LDOS in the presence of small disorder averaged over four disorder realizations. The system parameters for \textbf{c} – \textbf{d} are $\ell = 18a$, $\Delta = 0.4t$, $E_F = 2.2t$, $JS = 0.8t$. \textbf{e} The lowest positive energy eigenstate confined to the edge in the pristine system. \textbf{f} Edge mode LDOS averaged over three disorder realizations. The system parameters for \textbf{e} – \textbf{f} are $l = 18a$, $\Delta = 0.4t$, $E_F = 2.2t$, $JS = 0.8t$.
	}
\end{figure}

\begin{figure}[tbp]
	\centering
	\includegraphics[width=0.99\linewidth]{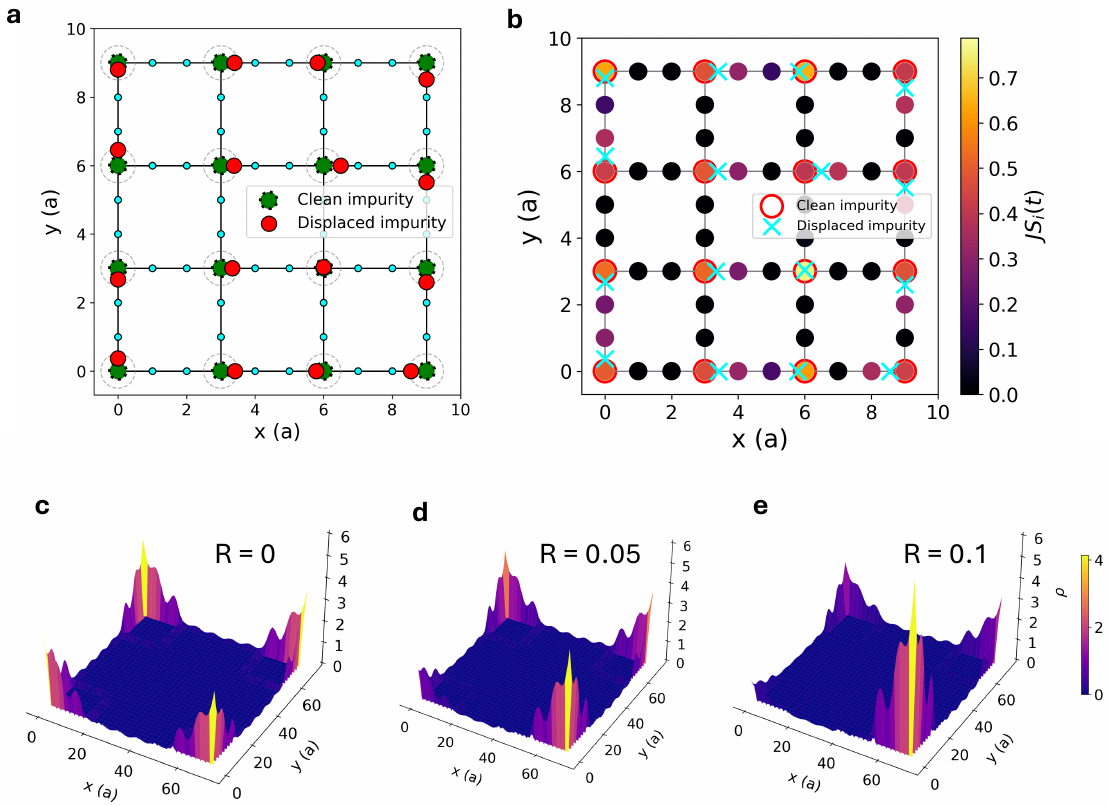} 
	\caption{\label{fig.r11} \textbf{a} A schematic model for the random displacement of magnetic impurities from the intersections within a disk of radius $R$. \textbf{b} A schematic for the $JS$ profile of the disordered networks. \textbf{c} The LDOS for a pristine network with $\ell = 18a$ (same parameters as those used in Fig.~{\ref{fig.r09}}). \textbf{d}-\textbf{e} The corner modes with small random displacement of the magnetic impurities for two disorder configurations, $R = 0.05 a$ and $R = 0.1 a$, respectively.}
\end{figure}

\begin{figure}[tbp]
	\centering
	\includegraphics[width=0.99\linewidth]{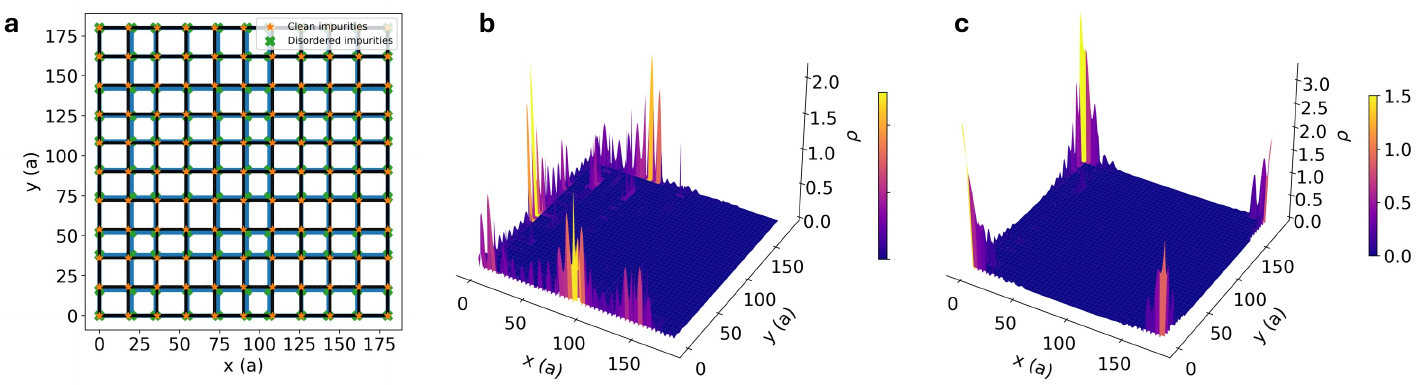} 
	\caption{\label{fig.r12} \textbf{a} A schematic model for the wire shift disorder. \textbf{b} The effect of shifts on the edge modes averaged over four different disorder realizations. \textbf{a} Corner modes averaged over six disorder realizations. The parameters for these calculations are: $\ell = 18a$, $\Delta = 0.4t$, $E_F = 2.2t$, $JS = 0.8t$.
	}
\end{figure}

\begin{figure}[tbp]
	\centering
	\includegraphics[width=0.99\linewidth]{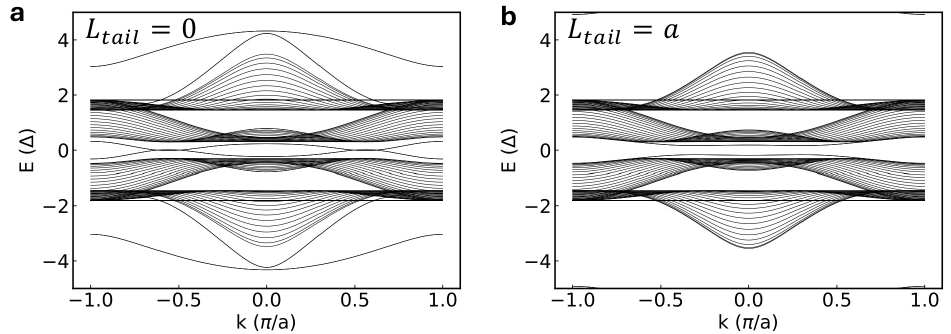} 
	\caption{\label{fig.r13}
    \textbf{a} Edge dispersion corresponding to the parameters used to generate Fig. 2c in the main text.
    \textbf{b} Edge dispersion corresponding to the parameters used to generate Fig. 2c in the main text with a tail of length $L_{tail} = a$ appended to the boundary at the vertices hosting magnetic impurities.
	}
\end{figure}

\section{Effects of disorder}
We present a three types of disorder calculations and show the boundary modes in bulk-dissociated phases are resilient to weak structural disorder that break the discrete rotational symmetry of the network. In Fig.~{\ref{fig.r09}}a-b, we provide a schematic illustration of a disorder implementation for the magnetic impurity sites where we allow for the magnetic impurity sites at the vertices to be randomly displaced to another site in the network within a radius $R$ of the vertex. 
The random displacement was sampled within a disc of radius $R$ using the relation $r = R\sqrt{u}$ where u represents a uniform random number in the interval $u \in [0,1)$ . Since the impurities can only be displaced along the horizontal or vertical wire, we apply constraints on the angle of the position vector r from the vertices so that the displacement, $\Delta r \in ( r \cos \theta, r \sin \theta)$ with $\theta =0, \pi, \pi/2, 3 \pi/2$. The exchange term takes a fixed value, $JS$.
To analyze the effect on corner modes, we evaluate the local density of states (LDOS) in a pristine network (Fig.~{\ref{fig.r09}}c) and compare it to the LDOS averaged over four disorder realizations with $R=0.56a$ (Fig.~{\ref{fig.r09}}d). In Fig.~{\ref{fig.r09}}d, we see the random displacements break the $C_4$ symmetry of the LDOS profile while keeping the corner modes intact. Similar calculations for bulk-dissociated edge modes are shown in Fig.~{\ref{fig.r09}}e-f. Here we also see the $C_4$ symmetry of the LDOS is strongly broken in Fig.~{\ref{fig.r09}}f, but the average over disorder realizations shows that the edge modes are resilient to small random displacements of the magnetic impurities.

In Fig.~{\ref{fig.r11}}, we present disorder in the amplitude of $JS$ at vertices of the network. 
The random sampling of the impurity position is the same as before but now the exchange term is applied to the single two nearest sites of the displaced impurity position with a linear distribution such that the sum of the exchange terms on the two sites is $JS$ i.e. for a random impurity are $\bf r_i$, the two nearest sites $\bf r_{i,1}$ and $\bf r_{i,2}$ have exchange terms $\frac{JS}{a} \vert {\bf r_i - r_{i,1}} \vert$ and $\frac{JS}{a} \vert {\bf r_i - r_{i,2}} \vert$, respectively. 
In Fig.~{\ref{fig.r11}}c-e, we present similar disorder calculations to those shown in Fig.~{\ref{fig.r09}}c-e and find that corner eigenstates have slightly broken $C_4$ symmetry.

In Fig.~{\ref{fig.r12}}, we present the impact of random displacements of the network wires. 
The random shifts of the wires along x and y are sampled with discrete random numbers: $\Delta x_i, \Delta y_i \in [-W_s, W_s]$, where $W_s$ is the maximum wire shift. The wires at the boundary are kept fixed while the inner wires are shifted randomly. The impurity positions are the new intersections of the wires resulting from the random shifts of the wires. 
In Fig.~{\ref{fig.r12}}b-c, we present the disorder-averaged LDOS of edge and corner modes with broken $C_4$ symmetry with $W_s =2a$. Here, we see both boundary modes remain confined to the boundary. We also analyzed the impact of adding a tail to the boundary of the network. 
In Fig.~{\ref{fig.r13}}a, we show the dispersion of phase (II) in Fig. 2 of the manuscript and 
Fig.~{\ref{fig.r13}}b shows how the bulk-dissociated edge band merge with the bulk bands 
with tails on the boundary.

\begin{figure}[tbp]
	\centering
	\includegraphics[width=0.99\linewidth]{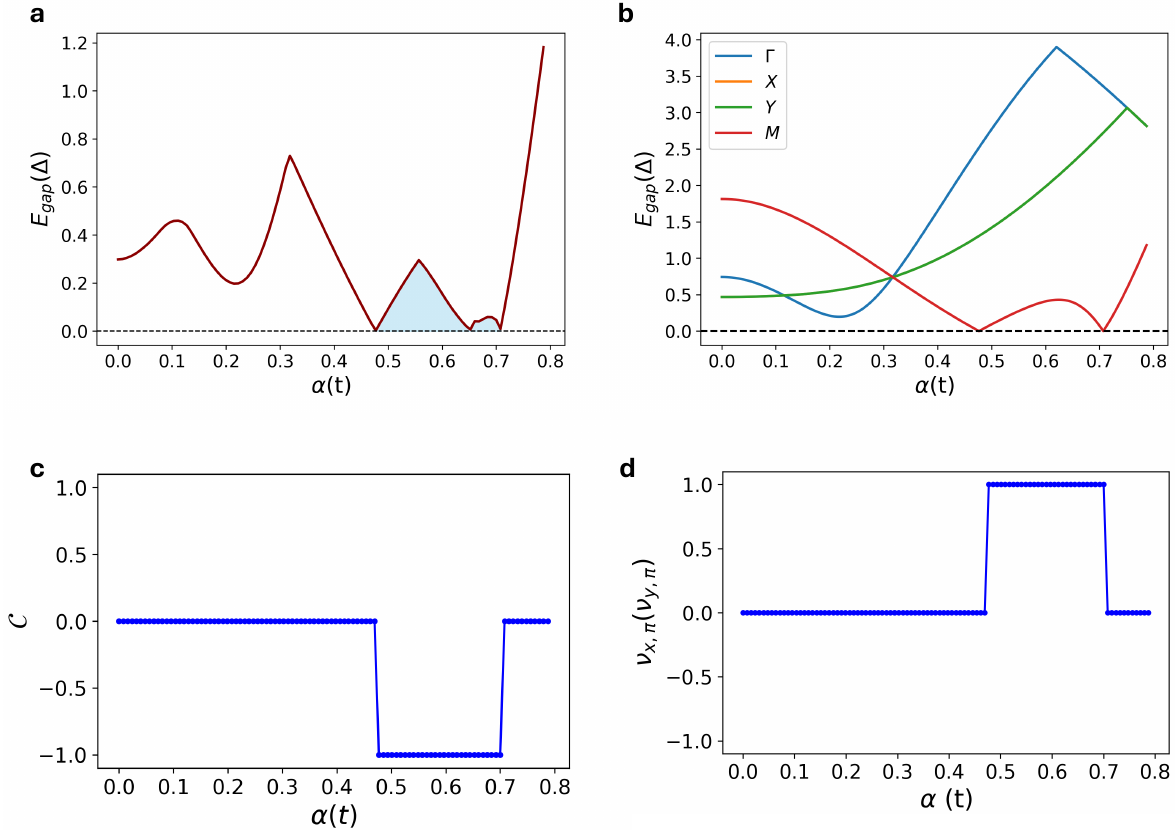} 
	\caption{\label{fig.r01} Bulk gap, strong and weak topological invariants. \textbf{a} $\ell = 10a$, $\Delta = 0.4t$, $E_F = 2.15t$, $JS = 0.8t$. The system is in the strong and weak topological phases in the range $ 0.5t < \alpha < 0.7t$ as indicated by a Chern number of -1 (panel \textbf{c}) and a weak index of +1 (panel \textbf{d}). The topological phase transition coincides with the bulk gap closing at the high symmetry M point (panel \textbf{b}).
	}
\end{figure}

\section{Effect of Rashba spin-orbit coupling}
In Fig.~{\ref{fig.r01}}, we present a calculation of the bulk gap and topological invariants as a function of Rashba spin-orbit coupling, 
$
h_R = \alpha a\left({\bf k \times \sigma} \right)_z.
$
In Fig.~{\ref{fig.r01}}a, we plot the spectral gap of a YSR network versus $\alpha$ in where the network is in the bulk-dissociated phase at $\alpha = 0$. We see a gap closing occurs near $\alpha = 0.5t$, which occurs at the M point of the Brillouin zone shown in Fig.~{\ref{fig.r01}}b. In Fig.~{\ref{fig.r01}}c-d, we see the gap closing coincides to the onset of a non-zero Chern number and weak invariants. Thus, strong topological superconductivity can be induced in YSR networks with Rashba spin-orbit interactions.

\begin{figure}[tbp]
	\centering
	\includegraphics[width=0.99\linewidth]{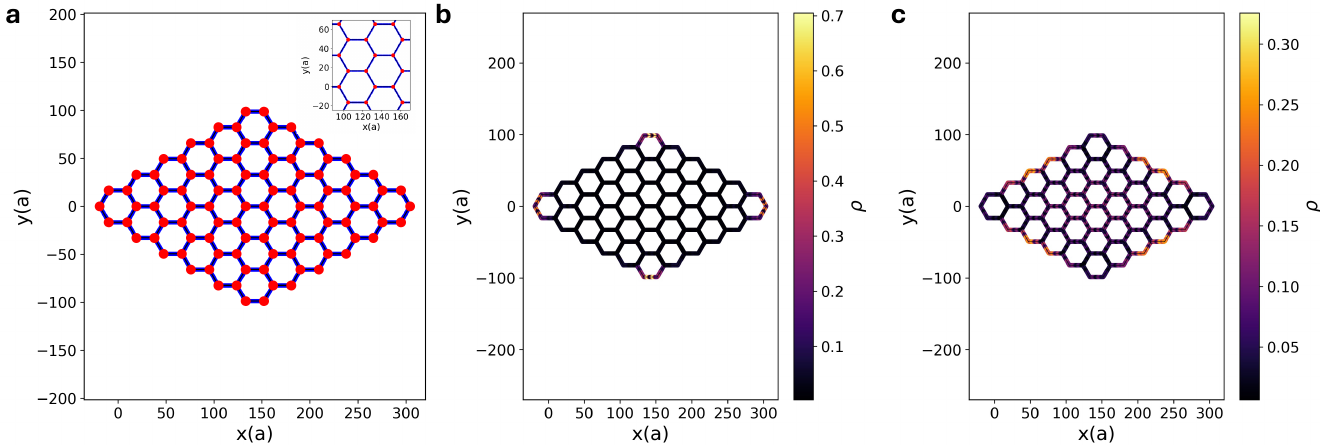} 
	\caption{\label{fig.r17} \textbf{a} Network lattice for the honeycomb YSR network model. The honeycomb YSR network is modeled in a similar way as the square YSR network, i.e., the magnetic impurities sit at each vortex of the hexagon, and there are $\ell$ lattice sites with lattice constant $a$ between neighboring vertices of the hexagon. The LDOS of \textbf{b} corner modes \textbf{c} edge modes.
	}
\end{figure}

\section{Honeycomb network geometry}
We performed a preliminary analysis of the honeycomb YSR network which is depicted in the schematic of decorated MATBG. In Fig.~{\ref{fig.r17}}a, we show the lattice model used to construct a honeycomb network with zig-zag boundaries. 
The primitive translational vectors are ${\bf a_1} = \ell ( 3/2, \sqrt{3}/2)$ and ${\bf a_2} = \ell ( 3/2, -\sqrt{3}/2)$, where $\ell = (n+1) a$,  the distance between the two vertices.
The center of each hexagon is  ${\bf R_{ij}} = i {\bf a_1} + j {\bf a_2}$ where $i = 0, ..., (N_x^{hex} – 1)$ and $j = 0, ..., (N_y^{hex} – 1)$. $N_x^{hex}$ and $N_y^{hex}$ are the number of hexagons along the x and y directions. The six vertices of the hexagon are generated with the angles $\theta_n =  n\pi/3,~ n=0,1,...,5$. The vertex positions relative to the center of the hexagon are given by $v_n= \ell (\cos\theta_n, \sin\theta_n)$.
In Fig.~{\ref{fig.r17}}b-c, we show the LDOS of corner and weak edge bound states of the network. This suggests the results in square network model we focused on in the original manuscript is relevant for other types of network geometries.

\bibliography{refs_Cuozzo}

\end{document}